\documentclass[12pt,preprint]{aastex}
\begin{document}

\title{Seoul National University Bright Quasar Survey in Optical (SNUQSO) II:  
      Discovery of 40 Bright Quasars near the Galactic Plane}

\author{Myungshin Im\altaffilmark{1,2}, Induk Lee\altaffilmark{1,3},
 Yunseok Cho\altaffilmark{1}, Changsu Choi\altaffilmark{1}, 
 Jongwan Ko\altaffilmark{1}, and Mimi Song\altaffilmark{1}}

\altaffiltext{1}{Astronomy Program, Department of Physics and Astronomy, FPRD, 
 Seoul National University, Seoul, Korea}
\altaffiltext{2}{mim@astro.snu.ac.kr}
\altaffiltext{3}{idlee@astro.snu.ac.kr}

\newcommand{\oiii}{[$\ion{O}{3}$]}
\newcommand{\hb}{$H\beta$}
\newcommand{\ha}{$H\alpha$}

\begin{abstract}
  We report the discovery of 40 bright quasars and active galactic nuclei
 (AGNs) at low Galactic latitude ($b < \vert 20\degr \vert$).
  The low galactic latitude region has been considered as a place
 to avoid when searching for extragalactic sources, 
 because of the high Galactic extinction as well as a large number of
 stars contaminating the sample selection.
  Bright quasars ($R \lesssim 17$) suffer more from such difficulties
  because they look like bright stars, which are numerous at low $b$, yet their
 surface number density is very low. 
  In order to find quasars in this region of the sky less explored  
 for extragalactic sources, we have started a survey of low Galactic
 latitude bright quasars as a part of the Seoul National University Quasar
 Survey in Optical (SNUQSO).
  Quasar candidates have been selected from radio and near-infrared (NIR) data.
  Out of 88 targets, we identify 29 bright quasars/AGNs around
 the antigalactic center, and 11 bright quasars/AGNs in the outskirts
 of the Galactic center, from two observing runs in 2006 at 
 the Bohyunsan Optical Astronomical Observatory (BOAO) in Korea
  Our finding demonstrates that quasars/AGNs can be discovered effectively
 even at low Galactic latitude using multi-wavelength data. 
\end{abstract}


\section{INTRODUCTION}

  Since the discovery of quasar 3C273 in 1963 (Schmidt 1963),
 more than 100,000 quasars and active galactic nuclei have been found 
 to date (Schneider et al. 2005; Croom et al. 2004; 
 Hagen  et al. 1995; Hewett et al. 1995;
 Becker et al. 2001; Schmidt \& Green 1983;
 Veron-Cetty \& Veron 2006).
  Despite of the rapid increase in the number of known quasars,
 however, only a small number of quasars have been discovered
 near the Galactic plane.
  The dots and crosses in Figure. 1 represent the positions of quasars
 discovered to date, taken from Veron-Cetty \& Veron (2006).
  Clearly, there is a lack of quasars/AGNs at low Galactic latitude
 ($b \lesssim \vert 20 \vert$).
  There are two reasons for this paucity of quasars at low Galactic latitude.
  First, there are simply too many stars that look like quasars 
 at low Galactic latitude so that it is much more challenging to select quasar 
 candidates in such a region than at a high Galactic latitude region.
  High Galactic extinction in the Galactic plane also causes
 a trouble when searching for quasars. Not only does Galactic extinction
 dim the light from quasars by several magnitudes or more, but it also 
 affects the optical colors of quasars, making it 
 difficult to isolate quasar candidates with otherwise highly efficient
 quasar selection methods employing optical color-color diagrams
 (e.g., Richards et al. 2004).

  Among quasars, bright quasars ($V \lesssim 17$ mag) get special attention.
  Because of their brightness, they are easy to use 
 to investigate the intergalactic medium
 (e.g., Weymann et al. 1981;
 Rauch 1998;  Fan et al. 2006), the properties of supermassive
 blackholes and their host galaxies (e.g., Kaspi et al. 2000;
 Kim et al. 2006),
 and the physical state and distribution of gas in our Galaxy
 (Sternberg 2003; Sembach et al. 2003; Savage \& Sembach 1996). 
  Because of their usefulness,
  bright quasars have been searched for extensively in 
 high Galactic latitude regions, especially in the northern hemisphere
 (Schmidt \& Green 1983; Hagen  et al. 1995; Hewett et al. 1995;
  Becker et al. 2001; Schneider et al. 2005).
  These surveys appear to have found the majority of bright quasars
 in their respective search areas, although a small number of bright quasars
 are still waiting to be discovered (Lee et al. 2008).

   Bright quasars at low Galactic latitude also have many 
 astrophysical applications. They can allow us to study 
 the gaseous component of matter in our galaxy from a unique point of view
 through their absorption lines, and offer an opportunity to study 
 their host galaxies using bright stars close to them as  guide stars for
 adaptive optics. They can also serve as reference points to examine 
 the motion of stars in our Galaxy.
  However, finding bright quasars at low Galactic latitude 
 seems to be a daunting job considering the aforementioned difficulties.
  A simple quantitative argument is given below to demonstrate how 
 difficult it is to identify bright quasars at low Galactic latitude.
  The surface number density of bright quasars at $V < 17$ mag is
 0.1 $deg^{-2}$, while the number density of stars at low Galactic 
 latitude ranges from a few thousand (anti-galactic center) to 
 a few tens of thousand per square degree at around $l=0^{\circ}$
 and $b=10^{\circ}$ (Robin et al.2003). This is 10 to 100 times more than
 the number density of stars at high Galactic latitude (Spagna et al. 1996).
  These numbers become even worse if we take into account Galactic
 extinction which dims the light from extragalactic
 objects by more than a magnitude in most areas at low Galactic latitude.
  Assuming 1 magnitude extinction,
 the apparent magnitude cut of $V < 17$  mag at low Galactic latitude
 would actually sample quasars with $V < 16$ mag if they were at high Galactic
 latitude.  
  At $V < 17$ mag, the number density of quasars is only 0.01 $deg^{-2}$.
  Therefore, the quasar-to-star ratio of an apparent magnitude-limited sample
 at $V < 17$ mag becomes 1 in a few hundred thousand to a few million,
 or 100 to 1000 times worse than at high Galactic latitude.  
  At $V \lesssim 17$, the optical multi-color selection gives a 
 quasar identification efficiency of about 90\%, but at  low Galactic
 latitude this efficiency would drop to 9\% to 0.9\% following the above 
 argument. Clearly an alternative
 method is needed for selecting bright quasars at low Galactic latitude.

   In order to broaden our knowledge of bright quasars toward   
 the zone of avoidance, we have started a survey of bright quasars
 at low Galactic latitude regions, as a part of the Seoul National University
 Quasar Survey in Optical (SNUQSO). Here, we report the discovery of
 40 bright quasars/AGNs at low Galactic latitude, using a selection
 algorithm designed to efficiently sample low Galactic latitude quasars.
   Throughout the paper, the magnitudes are Vega-based, unless otherwise
 mentioned. Also, we adopt the cosmological parameters of $\Omega_{m} = 0.3$, 
 $\Lambda = 0.7$, and $H_{0} = 70~km\,s^{-1}\,Mpc^{-1}$. 

\section{SELECTION OF CANDIDATES}

\subsection{Method}

    The key to successfully finding low Galactic latitude quasars lies in   
  an efficient algorithm that can distinguish quasars from 
  the plethora of stars and dust in the Galactic plane.
   To achieve such a goal, we use a combination of radio 
  and near-infrared (NIR) data. For the radio data,
  we use the NRAO VLA Sky Survey (NVSS; Condon et al. 1998) data set,
  and for the NIR data, we use the Two Micron All Sky Survey (2MASS) data set
  (Skrutskie et al. 2006).
   These data sets are chosen because they provide nearly complete coverage
  of the Galactic plane to the depth sufficient for finding a fair number
  of bright quasars. The NVSS data covers the whole sky above 
  $\delta = -40^{\circ}$  to the depth of 2.5 mJy in 21cm. The 2MASS data
  covers the entire sky to the depth of $K = 15$ mag.
   The reason for the use of the radio and NIR data sets is elaborated 
 below.

   Our selection of candidates starts with identifying radio sources at
 $\vert b \vert < 20$.  As the name ``quasar'' originates from the word  
 ``quasi-stellar radio star,'' it has been widely recognized that
 radio emitting point sources are very likely to be quasars or AGNs. 
   The usual procedure for finding quasars using radio data is to find 
 an optical counterpart to the radio source and perform  spectroscopic
 observation to determine whether it is a quasar.
  In high Galactic latitude 
 regions, it is not very difficult to identify the optical counterpart
 of a radio source, given that the radio data have a reasonably good 
 positional accuracy. For example, the FIRST survey data show that
  the matching radius of 1\arcsec~ is sufficiently good for finding an optical
  counterpart (White et al. 2000).
  However, that approach is not directly applicable in this case. 
  The beam size of the NVSS data is large (FWHM = 45\arcsec)
 compared to FIRST, and the quoted positional accuracy of point sources
 varies from $rms \sim 1\arcsec$ (for sources brighter than 15 mJy)
 to 7 \arcsec ~(at the survey limit of 2.5 mJy), forcing us 
 to adopt a conservative 2-$\sigma$ radius of 15\arcsec~  
 as a search radius to identify the optical counterpart.
 However, within a 15\arcsec~ radius circle of a NVSS radio source 
at low Galactic latitude, there are on average $\sim$6 point
 sources down to the optical magnitude of  $R \lesssim 19$ mag  
 and sometimes more. 

  In order to improve the quasar selection efficiency by weeding out
 stars near the radio position,  we add a NIR selection method.
  Low redshift quasars are known to have
 red $J-K$ colors due to a steeply rising NIR spectra
 (e.g., Glikman et al. 2006), and NIR colors 
 have been used previously to select quasars and AGNs (e.g., Figure 5 in
 Barkhouse \& Hall 2001; see also, Nelson et al. 1998;
 Cutri et al. 2000; Maddox \& Hewett 2006). Therefore, we applied a NIR
 color cut on point sources near the radio position and selected 
  objects with  $J-K > 1.4$ as quasar candidates.
  The NIR cut of $J-K > 1.4$ has been found to be the optimal 
 cut for minimizing the contamination from stars and galaxies by 
 performing the above selection method on known quasars/galaxies/stars
 in the fourth data release of the Sloan Digital Sky Survey (SDSS; 
 Adelman-McCarthy, et al. 2006).
 In general, the selected candidates have $K < 14$ mag 
 and an optical magnitude of $R < 19$.
 With the NIR cut, the contamination rate of stars/galaxies
 on the SDSS sample with the above method is found to be on the order
 of a few percent or less, but at the expense of preferentially selecting
 low redshift quasars at $z \lesssim 0.4$ (see also, Chiu et al. 2007).
  A more complete description of the sample selection and its 
 efficiency will appear in I. Lee et al. (2008, in preparation).

  Although our method can be applied to a very crowded region in 
 the Galactic plane, we have limited the initial survey area to the following
 region so that we get a reasonably high quasar identification efficiency
 of $\sim 50$\%.

\[ \begin{array}{ll}
  \vert b \vert < 20^{\circ}  &  \mbox{at $120^{\circ} < l < 240^{\circ}$}\\
  5^{\circ} < \vert b \vert < 20^{\circ}  & \mbox{at $40^{\circ} < l < 120^{\circ}$ \& $240^{\circ} < l < 320^{\circ}$}\\
  10^{\circ} < \vert b \vert < 20^{\circ} & \mbox{at $0^{\circ} < l < 40^{\circ}$ \& $320^{\circ} < l < 360^{\circ}$}
 \end{array} 
\]

  Using this selection method, we identify 601 candidates
 in the above area covering 10919.9 $deg^{2}$, 88 of which were
 observed over $\sim$ 2700 $deg^{2}$ during 2006, as described
 in the next section.

\section{OBSERVATION}

  For spectroscopic confirmation of quasars, 
 we observed the quasar candidates
 using the longslit low-resolution spectrograph mounted
 on the Bohyunsan Observatory Eschelle Spectrograph (BOES) of
 the 1.8m telescope at  Bohyunsan Optical Astronomical Observatory
 (BOAO; Kim et al. 2003).
  The spectroscopic observations were carried out on 2006 June 17 - 24 and
 2006 December 20 - 25, among which about four nights were
 in observable condition under the seeing of $2$ -- $3$\arcsec.
  Using a 150 per mm grating and a slit width of 3.6\arcsec,
 we covered the wavelength range of 4200 - 8400 \AA~  at the spectral
 resolution of $\lambda/\delta\lambda \sim 366$, or $\delta v \sim 820
 ~km\, s^{-1}$. The observational setting is basically identical to 
 the default observational setting of SNUQSO, which is described in 
 more detail in Lee et al. (2008). 
  The June run targeted bright quasar candidates in the outskirts
 of the Galactic center, while in the December run, we observed targets near
 the anti-galactic center. 
  We observed 33 candidates in June, and 55 candidates in December.
  The observed data were reduced with IRAF\footnote{IRAF is
  distributed by the National Optical Astronomy Observatory, which
  is operated by the Association of Universities for Research in
  Astronomy, Inc., under cooperative agreement with the National
  Science Foundation.} packages, following a standard
 procedure including flux calibration. We estimate the accuracy of
 the relative flux calibration (as a function of 
 wavelength) to be roughly 15\% (Lee et al. 2007a).
  We originally took optical magnitudes from the USNO-A2.0 catalog 
 (Monet et al. 1998) for the quasar 
 candidates, but some of the apparent magnitudes appeared to be too bright.
  In order to check the USNO magnitudes, we performed the 
 multi-band imaging observation of objects that are identified as 
 quasars/AGNs using the 1.5m telescope at the Mt. Maidanak Observatory
 in Uzbekistan during 2006 August 1 - 12 and also in 2006 June/July 
 (M. Im et al. 2008, in preparation). The median seeing during the
 imaging run hovered around 0.7\arcsec to 1.0\arcsec, and the observations
 were carried out in photometric condition.
  Our imaging observation reveals that the USNO magnitudes are
 too bright by more than 1 mag for the four objects with
 USNO magnitude $R < 15$ (object numbers, 33, 36, 37, and 39 in 
 Table 1), and we have revised the $R$-band and $B$-band
 magnitude accordingly. In other cases, we find that 
 the average difference in $R$-band magnitude 
 is $R$(USNO) - $R$(Maidanak) $\simeq$ 0.12 mag.
  Table 1 lists the apparent magnitudes of the low Galactic
 latitude quasars/AGNs.
  The $R$-band and $B$-band magnitude of objects 29-39 in Table 1 are from
 our imaging observation, while the rest come from the USNO catalog.
  For some of the problematic cases, the blending of objects might be 
 responsible for the bright USNO magnitude since we find neighbors
 within $\sim$6\arcsec of the quasar.
  For the rest, it is not clear why the USNO magnitudes are so bright. 
  Monitoring of the quasars/AGNs may reveal the temporal variation of 
 brightness of such quasars/AGNs.
  Based on our imaging data of 11 objects, 
  we note that the magnitudes in Table 2 originating from the USNO
 catalog may be too bright when $R < 15$ mag.

\section{RESULTS AND DISCUSSIONS}

  From the observed spectra, quasars were identified as objects with
 redshifted, broad Balmer emission lines with their FWHM line widths 
 exceeding 1000 $km\,s^{-1}$ (White et al. 2000; Schneider et al. 2003).  
  With such a criterion, we find 40 quasars among 88 candidates. Application
 of an absolute magnitude cut of $M_{i} < -22$ AB magnitude
 (Schneider et al. 2003) reduces this number to 
 37\footnote{$M_{i}$ can be calculated from the $R$-band absolute magnitude 
 listed in Table 1, by applying the color correction of
 $R$(Vega) - $i_{AB}$ $\simeq$ -0.1 mag, which we derived using 
 a template quasar spectrum from Vanden Berk et al. (2001). Here, 
 $R$(Vega) - $i_{AB}$ would be $\sim$ -0.2 mag for the $R$-band 
 magnitudes from USNO based on the comparison of the USNO $R$-band magnitude 
 to our data.}

  Figure 1 shows the distribution of the 40 new quasars/AGNs in Galactic 
 coordinates (triangles and squares).
  Also plotted in the figure are quasars and active galaxies 
 listed in Veron-Cetty \& Veron (2006). We also plot quasars brighter
 than $V = 17$ mag as crosses.
  The lack of bright quasars at low Galactic latitude is apparent from
 this figure. Especially in the Galactic plane near the anti-galactic 
 center ($120^{\circ} < l < 240^{\circ}$
 and $\vert b \vert < 8^{\circ}$), only 7 bright 
 quasars/AGNs have been reported previously. We more than double 
 this number now by discovering 13 additional quasars/AGNs in this
 region.

  In Figure 2, we present the spectra of the 40 quasars/AGNs, while  
 Table 1 summarizes the basic observed properties of these quasars.
 The extinction corrected absolute $R$-band magnitude, $M_{R}$,
 is calculated with the formula,

\[M_{R} = m_{R} - 5 \times log(d_{L}(z)) - 25 - K_{R}(z) - A_{R}. \]

  Here, $m_{R}$ is the $R$-band apparent magnitude 
  before extinction correction (col. [8] in Table 1), $d_{L}$ is the
  luminosity distance, $K_{R}(z)$ is 
  the $K$-correction,\footnote{Note that
 the correction is calculated assuming the Bessel $R$-band filter
 response on the 4k x 4k CCD Camera installed on the 1.5m
 telescope at the Maidanak Observatory (Ko et al. 2007,
 in preparation).
 Application of this
 $K$-correction to the photographic $R$-band magnitude from USNO may 
 result in a difference on the order of $\sim$0.1 mag.} 
 derived from the composite spectrum of SDSS quasars 
 (Vanden Berk et al. 2001), and $A_{R}$ is the Galactic extinction 
  toward the position of the quasar/AGN taken from NASA/IPAC Extragalactic 
  Database, which is based on
  the reddening map of Schlegel et al. (1998).
   Figure 2 shows a clear detection of redshifted H$\alpha$, H$\beta$,
 and [OIII] lines.  Redshifts of these quasars/AGNs range from $z = 0.05$ 
 to $0.3$, and the rest-frame FWHMs are all measured to be well above 
 1000 $km\,s^{-1}$, ranging from 1300 to 8000 $km\,s^{-1}$, showing that 
 they belong in the category of quasars and active galaxies
 (White et al. 2000; Schneider et al. 2003). We also find that 
 13 are X-ray sources and 27 are infrared sources.
  In particular, SNUQSO J210931.9-353258 has been previously identified as
 a gamma-ray blazar source by Sowards-Emmerd et al. (2005),
 but without redshift identification.
  We now identify this object to be a quasar at $z=0.202$.

  The $E(B-V)$ values in the direction of quasars/AGNs 
 range from 0.1 to 0.9 with the median value at
 $E(B-V) = 0.214$.
  We examined whether there is a correlation between 
 the Balmer decrement and the $E(B-V)$ value. To do so, spectra are 
 arranged according to the order of increasing $E(B-V)$ value in Figure 2.
  The figure shows that there is a general tendency of weakening of \hb~  flux
 with respect to \ha~  flux as the $E(B-V)$ values increase.
 We have plotted the \hb~  to \ha~  ratio versus $E(B-V)$ and confirmed this
 general trend, but find a large scatter in the relation too.
 Detailed model-fitting of spectra and the observation during photometric
 nights may reduce the scatter, 
 but it may not help much, since the large scatter could indicate 
 that the line ratios of some of the quasars/AGNs do not reflect
 the simple case B recombination
 (e.g., Glickman et al. 2006; Soifer et al. 2004).
 The intrinsic scatter in this ratio due to dust absorption
 within quasars/AGNs and their host galaxies also complicates the interpretation. 

  The radio-to-optical ratios ($R^{*}$) of the newly discovered quasars/AGNs
 range between 0.1 and 282 with the median value at 1.9. The radio-to-optical 
 flux ratio is defined to be the ratio of flux at the rest-frame 
 6 cm to 0.44 $\mu$m (Stocke et al. 1992). 
  The 6cm flux is derived from the 21cm flux assuming a power-law
 of $f_{\nu} \sim  \nu^{-0.46}$ (Ho \& Ulvestad 2001), and applying 
 a corresponding power-law $K$-correction in the form of 
 $K(z) = -2.5\,(\alpha + 1)\,log(1+z)$ where $\alpha=-0.46$.
  The 0.44 $\mu$m 
 flux is derived from the $B$-band apparent magnitude in Table 1,
 corrected for 
 Galactic extinction (Schlegel et al. 1998), as well as for the
 $K$-correction derived from the SDSS quasar composite spectrum of 
 Vanden Berk et al. (2001).
  We find six radio-loud quasars ($R^{*} > 10$) which corresponds
 to 16\% of our sample, excluding three AGNs. This number is slightly higher
 than the often-quoted value of 10\% from more complete surveys
 at the magnitude range of our sample (e.g., Figure 18
 in Ivezic et al. 2002; Hewett et al. 2001). The difference appears to 
 be at least partially related to the radio selection of our sample.

  In Figure 3, we plot the $R$-band magnitude distribution of 
 the newly discovered quasars/AGNs, before and after applying 
 the extinction correction assuming the reddening map of
 Schlegel et al. (1998). Before applying the extinction correction,
 the distribution peaks at $R \sim 16$ mag, with a wide tail
 toward $R \sim 18$ mag, showing that our 
 sample completeness is skewed toward the brighter magnitude of
 $R < 16$ mag. 
  After applying the extinction correction,
 many $R > 16$ mag quasars/AGNs move toward the $R < 16$ mag area, making 
 our sample appear as if it were selected to be $R \lesssim 16$ mag.
  Note that the Galactic extinctions, $A_{R}$ and $A_{B}$, range from 
 0.22 to 2.57 mag and 0.25 to 4.04 mag, respectively,
 toward the quasars/AGNs. 

  Some of the quasars are found to be fairly bright. If USNO apparent
 magnitudes are accurate (see Section 3),
 SNUQSO J035657.03+425540.6 and SNUQSO J062355.21+001843.4 would have
 $R \lesssim 14$ after Galactic extinction correction, making them
 two of the brightest quasars discovered to date.

  The 40 quasars/AGNs are discovered over about 2700 $deg^{2}$ of the sky,
 and therefore their surface number density is $0.014$ $deg^{-2}$
 down to $R \lesssim 16$ mag. We compare this number to the surface
 density of optically selected quasars/AGNs at high Galactic latitude 
 from La Franca \& Cristiani (1997), and also to the radio selected
 quasars from White et al. (2000).
  Assuming $B-R \simeq 0.7$ mag for
 quasars (Becker et al. 2001), our effective magnitude cut of
 $R < 16$ mag can be converted to 
 $B \lesssim 17$ mag. At that magnitude limit, La Franca \& Cristiani
 (1997) find the surface number density of quasars to be roughly
 0.05 $deg^{-2}$ (See also Croom et al. 2004).
  On the other hand, White et al. (2000) find about 70 quasars at
 $R \lesssim 16$ over 2682 $deg^{2}$ 
 which gives the surface density of 0.026 $deg^{-2}$.
  Our surface number density is nearly one half of the radio-selected
 sample of White et al. (2000).
   The major reason for the discrepancy is our 
 preferential selection of lower redshift quasars/AGNs with the adopted
 $J-K$ cut, which leads to the exclusion of about 50\% of $R < 16$
 quasars at $z \gtrsim 0.4$
 (see Figure 8 in White et al. 2000). Also, a part of the discrepancy
 can be explained with the difference in the depth of the radio data
 (2.5 mJy versus our  1 mJy).
  The discrepancy in the surface number density from the optically
 selected sample can be explained by the radio selection as well as 
 the reasons mentioned above.
  In any case, we expect that our sample is close to 
 complete at the NVSS radio flux limit only for quasars at $R < 16$ and $z \lesssim 0.4$.

  Objects which are not classified as quasars are mostly Galactic 
 objects, but some have featureless spectra similar to BL Lac objects.
  These objects will be discussed in a future work.

\section{CONCLUSION}
  
  By combining the radio and NIR quasar selection algorithm, 
  we have derived an algorithm which is efficient for
 selecting quasars with a minimal contamination rate.
  We applied this algorithm to search for quasars in 
 low Galactic latitude regions which are crowded with stars. From 
 the first two observing runs in 2006, we discovered 40 bright quasars/AGNs
 out of 88 candidates.
  The newly discovered quasars/AGNs will be useful for various
 astronomical studies
 such as the distribution of Galactic matter and the check of reddening 
 maps, and the quasar host galaxy study using a nearby bright star
 as a guide star.

\acknowledgements
  This work was supported by grant R01-2005-000-10610-0 from the Basic 
 Research Program of the Korea Science and Engineering Foundation.
  This work uses a CCD camera purchased by a special research
 grant from the College of Natural Sciences, Seoul National University.
  We thank the staff at BOAO, especially Kang-Min Kim  
 Byeong-Cheol Lee, and Youngbom Jeon for their professional aid
 during our observing runs, and
 the staff and scientists at the Maidanak Observatory,
 especially Rivkat Karimov and Mansur Ibrahimov for their assistance
 during the imaging observation. We also thank Patrick Hall, Luis C. Ho, 
 Roc Cutri, and Alastair Edge for useful discussions.

 Facilities: BOAO:1.8m, Maidanak:1.5m

\clearpage
\begin{deluxetable}{rllrrrllllllrrrrc}
\tablecolumns{17}
\tabletypesize{\scriptsize}
\rotate
\setlength{\tabcolsep}{0.03in}
\tablewidth{8.0in}
\tablenum{1}
\tablecaption{Properties of the newly discovered quasars/AGNs.}
\tablehead{
  \colhead{ID} & 
  \colhead{SNUQSO Name} &
  \colhead{R.A.} &
  \colhead{Decl.} & 
  \colhead{{\it l}} & 
  \colhead{{\it b}} &
  \colhead{$z$}&
  \colhead{$R$} & 
  \colhead{$B$} &
  \colhead{$K$} & 
  \colhead{$J-K$} &
  \colhead{$M_{R}$}&
  \colhead{$F$}& 
  \colhead{Ext.}&
  \colhead{$A_B$}&
  \colhead{$A_R$}&
  \colhead{{\it Comm.}}\\
  \colhead{} &
  \colhead{} &
  \colhead{(J2000)} &
  \colhead{(J2000)} &
  \colhead{(deg)} &
  \colhead{(deg)} &
  \colhead{} &
  \colhead{(mag)} &
  \colhead{(mag)} &
  \colhead{(mag)} &
  \colhead{(mag)} &
  \colhead{(mag)} &
  \colhead{(mJy)} &
  \colhead{(mag)} &
  \colhead{(mag)} &
  \colhead{(mag)} &
  \colhead{} \\
  \colhead{(1)} &
  \colhead{(2)} &
  \colhead{(3)} &
  \colhead{(4)} &
  \colhead{(5)} &
  \colhead{(6)} &
  \colhead{(7)} &
  \colhead{(8)} &
  \colhead{(9)} &
  \colhead{(10)} &
  \colhead{(11)} &
  \colhead{(12)} &
  \colhead{(13)} &
  \colhead{(14)} &
  \colhead{(15)} &
  \colhead{(16)} &
  \colhead{(17)} 
}
  
\startdata
1&J010629.99+461935.1&01:06:29.99 &46:19:35.1&125.64242 &-16.46592 & 0.1243		&16.3  &18.0 &12.81 &1.78 		&-23.0		&10.8 & 0.12	&	0.43	&	0.33 &\\
2&J022540.43+583457.1&02:25:40.43 &58:34:57.1&134.96545 &-2.06997 & 0.0971 		&17.0  &18.2 &12.75 &1.61 		&-23.9 		&8.5 &0.95 	&	4.04	&	2.57&IR\\
3&J035657.03+425540.6&03:56:57.03 &42:55:40.6&155.29287 &-7.99082 & 0.0659	 	&15.0  &16.1 &11.08 &2.17 		&-23.9 		&14.7 &0.54 	&	2.20	&	1.42 &IR, X\\
4&J041239.56+582518.2&04:12:39.56 &58:25:18.2&146.79854 &5.20963 & 0.0675 		&16.0  & 17.1 &12.26 &1.69 		&-23.7 		&3.0 &0.82 	&	3.46	&	2.21&\\
5&J042533.12+344719.5&04:25:33.12 &34:47:19.5&165.05205 &-10.03833 & 0.0578 		&15.4  &16.6 &11.30 &2.17 		&-22.5 		&5.8 &0.30 	&	1.18	&	0.79&X\\
6&J045458.67+694945.6&04:54:58.67 &69:49:45.6&141.21285 &16.18279 & 0.1401 		&15.2  &15.8 &12.10 &2.23 		&-24.4 		&3.5 &0.14 	&	0.49	&	0.36&IR\\
7&J050322.13+592656.0&05:03:22.13 &59:26:56.0&150.38249 &10.78847 & 0.0963   	&17.3 &19.0 &12.18 &2.19 		&-20.9 			&4.7 &0.66 	&	2.75	&	1.76&IR\\
8&J051148.53+282127.4&05:11:48.53 &28:21:27.4&176.38036 &-6.50323 & 0.1331		&17.6  &19.3 &12.89 &1.80 		&-23.4 		&19.0 &0.73 	&	3.02	&	1.93&\\
9&J051832.42+031609.4&05:18:32.42 &03:16:09.4&198.84259 &-18.92777 & 0.0572 		&15.6  &17.1 &12.64 &1.86 		&-21.8 		&9.0 &0.11 	&	0.38	&	0.30&\\
10&J052807.06+054011.7&05:28:07.06 &05:40:11.7&197.92783 &-15.65704 & 0.1012		&16.5  &17.2 &12.71 &1.83 		&-22.9 		&28.1 &0.35	&	1.45	&	0.96&IR\\
11&J052901.65+105709.8&05:29:01.65 &10:57:09.8&193.36578 &-12.75781 & 0.0764 		&15.8  &16.6 &12.84 &1.56 		&-22.9 		&3.7 &0.35 	&	1.37	&	0.91&X\\
12&J053221.32+393922.1&05:32:21.32 &39:39:22.1&169.46582 &3.39566 & 0.0720 		&18.3  &17.8 &12.51 &1.55 		&-21.3 		&2.8 &0.72 	&	3.07	&	1.97&IR\\
13&J054706.34+442026.7&05:47:06.34 &44:20:26.7&166.90147 &8.18474 & 0.1190 		&17.0  &16.8 &13.23 &1.60 		&-22.7		&4.0 &0.34 	&	1.35	&	0.90 &IR\\
14&J054854.50+363250.6&05:48:54.50 &36:32:50.6&173.83374 &4.52368 & 0.0745 		&17.2  &19.1 &13.58 &1.54 		&-22.4 		&5.8 &0.68 	&	2.86	&	1.83&IR\\
15&J055447.25+662043.8&05:54:47.25 &66:20:43.8&147.27239 &19.28736 & 0.1859 		&14.5  &14.6 &11.41 &2.17 		&-25.8 		&5.6 &0.16 	&	0.59	&	0.43&X\\
16&J060040.11+000618.6&06:00:40.11 &00:06:18.6&206.97350 &-11.22760 & 0.1147 		&17.6  &19.0 &11.75 &2.19 		&-23.2 		&212.7 &0.74 &	3.20	&	2.04&IR\\
17&J061713.50+285138.3&06:17:13.50 &28:51:38.3&183.48294 &5.97938 & 0.1036 		&16.4  &17.3 &12.01 &2.04 		&-23.9 		&10.7 &0.68 	&	2.82	&	1.81&IR,X\\
18&J061717.30+430056.0&06:17:17.30 &43:00:56.0&170.68829 &12.38432 & 0.2067 		&17.1  &17.3 &13.36 &1.73 		&-23.5 		&20.9 &0.15 	&	0.54	&	0.40&IR\\
19&J062355.21+001843.4&06:23:55.21 &00:18:43.4&209.50422 &-5.97636 & 0.0940		&14.8  &15.4 &11.09 &1.92 		&-25.2 		&46.7 &0.65 	&	2.68	&	1.72 &IR\\
20&J063100.52+503046.2&06:31:00.52 &50:30:46.2&164.56062 &17.54972 & 0.1805 		&16.8  &16.7 &13.08 &1.92 		&-23.4 		&2.7 &0.13 	&	0.45	&	0.34&IR\\
21&J063652.52$-$081048.6&06:36:52.52 &$-$08:10:48.6&218.59605 &-6.93935 & 0.2328 	&15.7  &15.6 &13.23 &1.64		&-26.1  	&2.7&0.48 &	2.00	&	1.30&X\\
22&J064139.37+451002.0&06:41:39.37 &45:10:02.0&170.45824 &17.25334 & 0.2973 		&16.3  &16.4 &12.84 &2.08 		&-25.1 		&24.3 &0.11	&	0.39	&	0.30&X\\
23&J064410.86+354644.6&06:44:10.86 &35:46:44.6&179.66110 &14.10731 & 0.0772 		&14.3  &15.9 &11.83 &1.80 		&-23.9 		&53.7 &0.16 	&	0.60	&	0.43  &\\
24&J064856.41+352206.6&06:48:56.41 &35:22:06.6&180.44495 &14.83466 & 0.1193 		&15.8  &16.2 &12.90 &1.75 		&-23.4 		&4.1 &0.12 	&	0.42	&	0.32 &X\\
25&J071014.55+131535.2&07:10:14.55 &13:15:35.2&203.12061 &10.11274 & 0.1411 		&15.8  &16.2 &12.32 &2.11 		&-23.7 		&6.7 &0.08 	&	0.25	&	0.22&IR\\
26&J071215.61+153930.3&07:12:15.61 &15:39:30.3&201.12942 &11.58509 & 0.1656		&15.3  &16.1 &12.42 &1.79 		&-25.0 		&4.1 &0.23 	&	0.99	&	0.68&X\\
27&J071645.09+002749.8&07:16:45.09 &00:27:49.8&215.41248 &5.84219 & 0.0519 		&15.1  &16.3 &12.13 &1.56 		&-22.3 		&4.1 &0.21 	&	0.79	&	0.55&IR\\
28&J073939.54$-$094456.2&07:39:39.54&$-$09:44:56.2&227.16980 &6.08370 & 0.1007	&15.7 &16.1    &12.20 &2.11		&-23.2 			&15.2 &0.17	&	0.66	&	0.47 &IR\\
29&J170631.57$-$064638.6&17:06:31.57&$-$06:46:38.6&14.02261 &19.67993 & 0.1046 	&16.5  &17.8 &13.03 &1.66 		&-23.5 		&7.2 &0.55 &	2.34	&	1.45 &IR, X\\
30&J175925.57+171513.5&17:59:25.57 &17:15:13.5&43.04643 &19.05558 & 0.1454		&16.0  &16.8 &13.11 &1.81 		&-23.6  	&14.3 &0.10 	&	0.44	&	0.27 &X\\
31&J180508.94+124253.9&18:05:08.94 &12:42:53.9&39.30022 &15.93342 & 0.1707 		&16.7  &17.8 &13.44 &1.82 		&-23.6 		&12.8 &0.22 	&	0.97	&	0.60 &IR\\
32&J193013.80+341049.5&19:30:13.80 &34:10:49.5&67.57943 &7.59368 & 0.0616 			&15.9 &17.7 &11.81 &2.32 	&-21.9   	&3.9 &0.19 		&	0.84	&	0.52 &IR\\
33&J193347.16+325426.0&19:33:47.16 &32:54:26.0&66.79822 &6.34119 & 0.0565 			&14.5 &15.5 &11.34 &1.70 	&-23.3 		&4.0 &0.27 		&	1.16	&	0.72 &IR, X\\
34&J193521.19+531411.9&19:35:21.19 &53:14:11.9&85.40431 &15.29883 & 0.2478 		&15.8  &16.5 &13.15 &1.54 		&-25.1 		&10.4 &0.11 	&	0.50	&	0.31 &IR\\
35&J203332.03+214622.3&20:33:32.03 &21:46:22.3&64.38411 &-10.77715 & 0.1735 		&16.2  &17.7 &13.05 &1.74 		&-23.9 		&340.4 &0.14 	&	0.59	&	0.36&\\
36&J210931.88+353257.6&21:09:31.88 &35:32:57.6&80.28167 &-8.35288 & 0.2023 		&14.7  &15.7 &13.03 &1.49 		&-25.8 		&1195.7 &0.16 	&	0.61	&	0.38 &IR, X, GB\\
37&J212235.46+273846.5&21:22:35.46 &27:38:46.5&76.26527 &-15.75300 & 0.0909 		&16.0  &17.2 &13.23 &1.52 		&-22.6 		&5.6 &0.13 	&	0.59	&	0.36&IR\\
38&J212756.45+271905.8&21:27:56.45 &27:19:05.8&76.86721 &-16.84005 & 0.1974 		&16.3  &18.3 &12.64 &2.21 		&-24.1 		&6.4 &0.12 	&	0.50	&	0.31   &IR\\
39&J222719.04+400550.2&22:27:19.04 &40:05:50.2&95.18443 &-14.88447 & 0.0659 		&15.6  &16.9 &11.91 &2.19 		&-22.2 		&6.9 &0.15 	&	0.63	&	0.39&IR\\
40&J230329.93+453740.4&23:03:29.93 &45:37:40.4&103.90851 &-13.21915 & 0.3007		&16.5  &17.1 &13.27 &1.89 		&-25.6 		&5.0 &0.35 	&	1.40	&	0.93 &IR\\
\enddata        
\tablenotetext{(8) - (10)}{$R$, $B$, and $K$-band apparent magnitudes, not corrected for Galactic extinction.}
\tablenotetext{(11)}{$J-K$ color, not corrected for Galactic extinction.}
\tablenotetext{(12)}{$R$-band absolute magnitude, corrected for Galactic extinction}                                                                                                                                            
\tablenotetext{(13)}{Flux at $1.4 ~GHz$ from NVSS.}
\tablenotetext{(14)}{Reddening parameter, $E(B-V)$}
\tablenotetext{(15),(16)}{Extinction at $B$ and $R$-bands in magnitude unit, respectively. $B$-band extinction is
calculated using $A_{B} = 3.91\, E(B-V)$ for USNO magnitudes,
and $A_{B} = 4.3 \, E(B-V)$ for the Maidanak data. For $A_{R}$, we used
$A_{R} = 2.93\,E(B-V)$ for both cases, since the difference in the extinction correction 
 arising   from a small discrepancy in conversion factor is less than 0.01 mag 
 between USNO-II $R$ magnitude and the Maidanak $R$ magnitude.}
\tablenotetext{(17)}{Comments. `X' stands for X-ray source, and `GB' for Gamma-ray Blazar Candidate.}
\end{deluxetable}

\clearpage
\begin{figure}
\label{fig1}
\figurenum{1}
\plotone{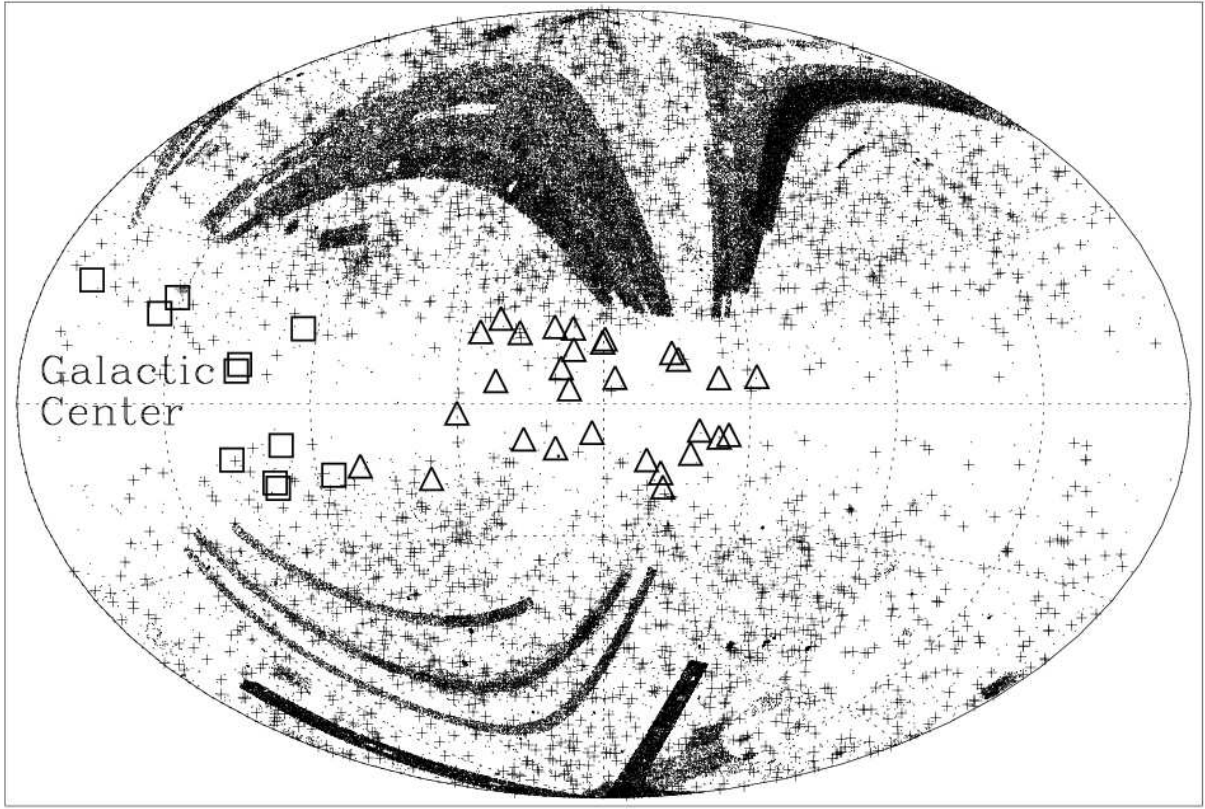}
\caption{Distribution of quasars and AGNs in the sky, plotted
 in Galactic coordinates. 
 Small dots represent the quasars/AGNs brighter than $V < 23$ mag 
 discovered to date, while crosses represent quasars/AGNs brighter than
 $V = 17$ mag.  Note a gap in the distribution 
 between the northern and southern Galactic hemispheres. The newly 
 discovered quasars are marked with squares (from 2006 June run) 
 and triangles (from 2006 December run). {\bf [High quality image is available on electronic journal]}
}
\end{figure}

\clearpage
\begin{figure}
\label{fig2}
\figurenum{2}
\plotone{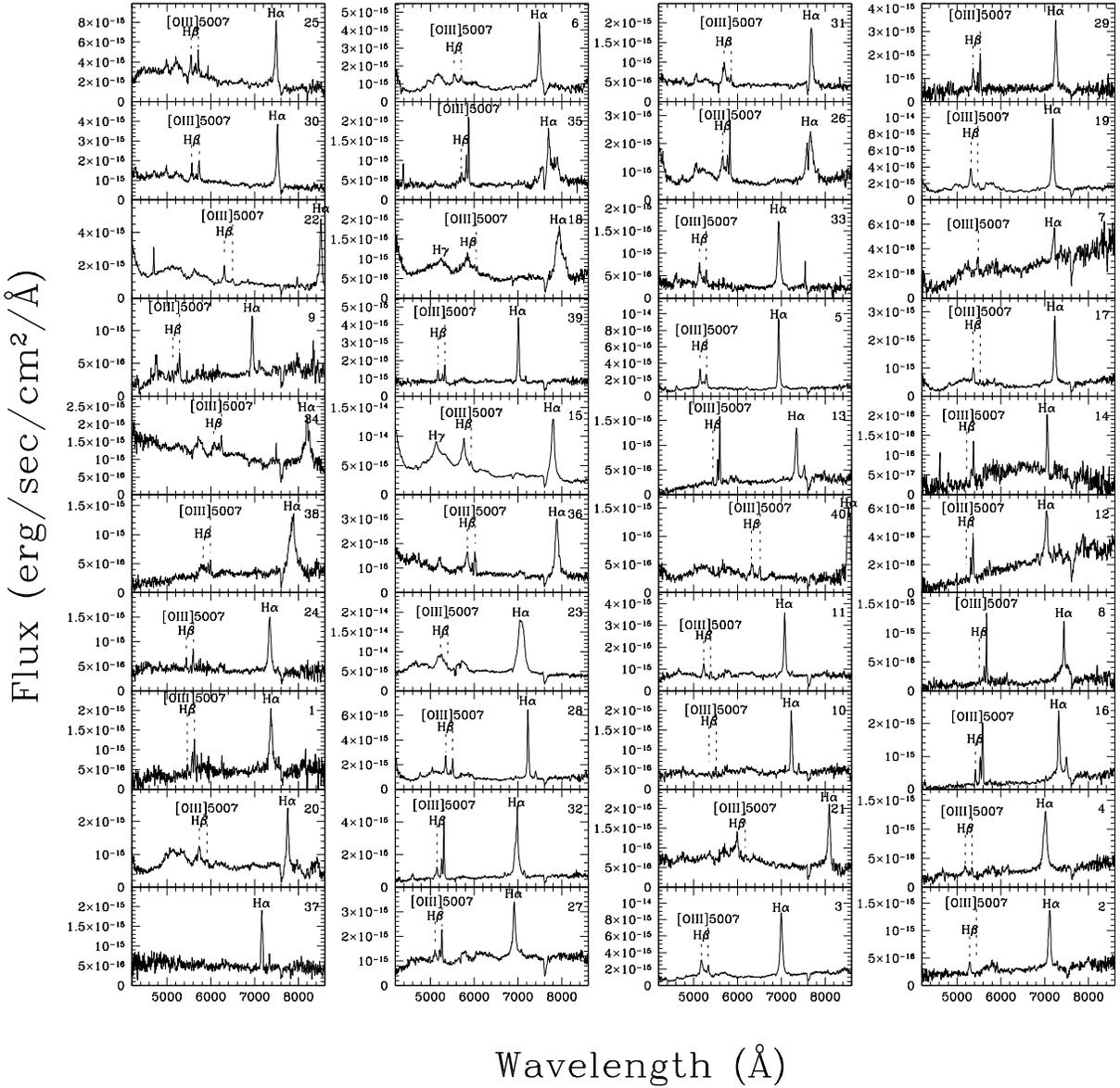}
\caption{
  Optical spectra of the newly discovered quasars, plotted in the order of 
 increasing $E(B-V)$ values. The $E(B-V)$ values increase 
 from top to bottom, left to right.
 Broad Balmer emission lines and $\oiii$ lines are visible
 for most of the quasars except for
 those at the location where the Galactic extinction is high.
}
\end{figure}

\clearpage
\begin{figure}
\label{fig3}
\figurenum{3}
\plotone{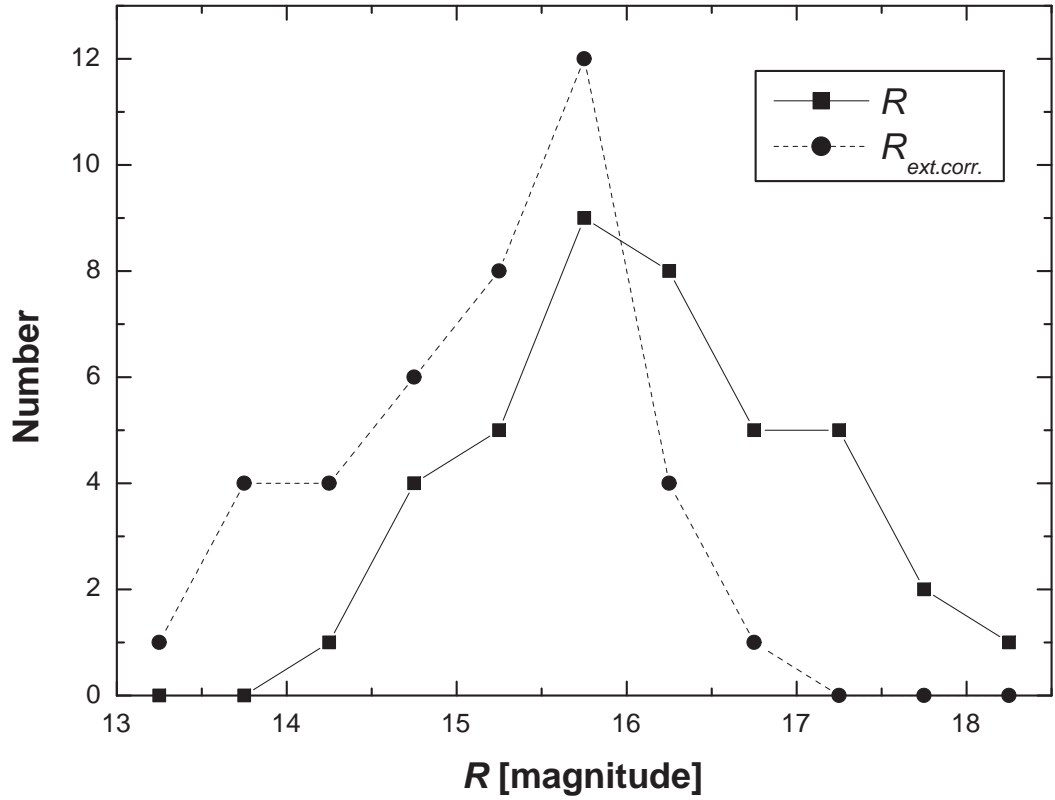}
\caption{
 $R$-band apparent magnitude of the newly discovered quasars 
 before (squares and solid line) and after (circles and dashed line)
 the extinction correction. From this figure and
 the analysis of the surface number density (see text), we find that
 our sample has an effective optical magnitude cut of $R < 16$ mag.
} 
\end{figure}

\end{document}